\documentclass[smallextended]{svjour3}
\usepackage{hyperref}
\hypersetup{
	colorlinks=true,
	urlcolor=blue,
	citecolor=blue}
\usepackage{pbox}
\usepackage[all]{xy,xypic}
\usepackage{amsfonts,amsmath,amsgen,amsopn,amsbsy,theorem,epsfig}
\usepackage{eufrak,amscd,bezier,latexsym,mathrsfs,enumerate}\usepackage[utf8]{inputenc}\usepackage[english]{babel}
\usepackage[dvipsnames]{xcolor}
\usepackage[pagewise]{lineno}
\usepackage{graphicx}
\usepackage{amssymb}
\usepackage{comment}
\begin{document}

%\begin{frontmatter}



\title{Towards a Blockchain based digital identity verification, record attestation and record sharing system}
 
\titlerunning{Towards a Blockchain based digital identity verification}        % if too long for running head

\author{Mehmet Aydar        \and
        Serkan Ayvaz  \and
        Salih Cemil Çetin
}

\institute{Mehmet Aydar \at
              AI Enablement Department, Huawei Turkey Research and Development Center, Istanbul, Turkey \\ 
              \email{maydar@kent.edu}             \\
           \and
           Serkan Ayvaz \at
              Department of Software Engineering,
              Bahcesehir University,
              Istanbul, Turkey \\ 
              \email{serkan.ayvaz@eng.bau.edu.tr}
          \and
           Salih Cemil Çetin \at
              AI Enablement Department, Huawei Turkey Research and Development Center, Istanbul, Turkey \\ 
              \email{salihcemil@gmail.com}      %  \\
}

\date{Received: date / Accepted: date}


% \author[label1,label3]{Mehmet Aydar}
% \address[label1]{Technology Introduction Department, Huawei Turkey Research and Development Center, %Inkılap Mah., Sanayi Cad. Onur Ofis Park, 34768  Umraniye,
% Istanbul, Turkey}
% \address[label3]{Department of Computer Science, Kent State University, Kent, OH, USA}
 
%  \author[label2]{Serkan Ayvaz}
% \address[label2]{Department of Software Engineering, Bahcesehir University, %Besiktas,  
% Istanbul, Turkey}
 
%\address[label2]{Address Two\fnref{label4}}

% \cortext[cor1]{Mehmet Aydar}
%\fntext[label3]{I also want to inform about\ldots}
%\fntext[label4]{Istanbul}

%\ead{maydar@kent.edu}

\maketitle

\begin{abstract}
The Covid-19 pandemic has made individuals and organizations to rethink the way of handling identity verification and credentials sharing particularly in quarantined situations. 
In this study, we investigate the inefficiencies of traditional identity systems, and discuss how a proper implementation of Blockchain technology would result in safer, more secure, privacy respecting and remote friendly identity systems. %are reviewed, and the potential benefits of employing Blockchain technology for a more efficient identity management system are explored. 
As a result, we propose a Blockchain based framework for digital identity verification, record attestation and record sharing, and we explain the framework in details with certain use cases. Our proposed framework promotes individuals to fully control their identity data and govern the level of the identity data sharing.
%When compared to traditional identity systems, it makes identity verification and identity based record sharing more efficient and secure, while respecting privacy of identity owners. By utilizing the trust fabric of Blockchain, the identity management system eliminates the middle man and waiting lines for authentication, authorization, and attestation. 
%maydar: framework oneriyoruz. pandemi gosterdiki online identity verificatio. remote signing. merkezi otoritelerin domine etmedigi fakat katkida bulundugu. uzaktan yeni kurumlara kyc. mevcut sistemleri inceledik. boyle olmasina karar verdik. use caseler kisa kisa.
\end{abstract}


%\end{frontmatter}


\section{Introduction}
\label{introduction}
Individuals identify themselves using various identity assets such as their name, national identity number and passport number. Identity assets are recorded in physical documents which are attested by central authorities. In the world of the internet, identity owners are required to provide their identity assets to institutions in order to verify their identities. Traditionally, institutions keep these sensitive information in centralized data silos. This traditional way of identity management methods are prone to data breaches, identity theft and fraud. Moreover, these methods have inefficiencies in terms of security, usability, privacy and globalization. As a matter of the fact, the digital transformation has further emphasized the need to move away from intermediary and provider-controlled identity management models toward user-controlled digital identity.

The advent of Blockchain technology has created an opportunity to transform how relationships between people and institutions are established and maintained. Blockchain technology can deliver secure solutions by integrating trust in the network itself. As for digital identity management, Blockchain technology can enable identity owners to have sovereignty of their identity based personal records, control access to their records and allow identity owners to share minimum amount of information while ensuring data integrity and trust. This study focuses on using Blockchain technology for identity management systems. 

The paper is organized as follows: Section \ref{problems} expresses the motivation of the study by describing current problems in traditional identity management methods. Section \ref{concepts} presents Blockchain technology, the main concepts used in Blockchain based identity management systems, and explains why the technology is suitable for a robust digital identity management system. In section \ref{proposed_solution}, we describe the proposed solution in details. Section \ref{related_work} reviews existing solutions in the field. Then, it is followed by conclusion.

\section{Problems in traditional identity management methods}
\label{problems}

The main challenges persisting in traditional identity management methods can be grouped into four categories: Usability, Privacy, Security and Globalization.

\subsection{Usability}
%maydar
%lack of remote identity verification,
%lack of remote signing
%lack of remote credential sharing
%lack of endless, multiple KYCs
%paper based documents, which are not healthy
Identity verification is challenging to handle remotely using traditional identity systems. During the pandemic caused by Covid-19 \cite{WHO_Pandemics}, a considerable amount of businesses moved their services online. As a result, remotely managing identity verification, Know Your Customer (KYC) requirements, document signing and credentials sharing became more essential for individuals, businesses and organizations. 

Identity owners typically use a combination of username/password in order to be authenticated for online services, in which they are required to provide their personal information in order to create an account. This leads to hassles such as having too many login information for each service, trying to remember various login credentials, and giving out private information for recovery of an account in case of forgetting the password. For remote authentication, identity owners are in most cases obliged to answer security questions containing their personal and sensitive information for verification of their identity. As a result, users pay a price by spending a great deal of time proving their identity, and risking their privacy by providing private identity information. A survey report by Centrify in 2014 indicated that even small businesses with 500 employees approximately lose \$200,000 annually in productivity due to the time spent on password management \cite{Centrify}.

\subsection{Privacy}  
%Only pertinent identify information is needed but we expose more: For instance for airtravel a full identification is required. In a nightclub they just need to see your age, for traffic stop they need to see your driving privileges. 
Identity assets define an individual. Yet, in traditional systems, individual's identity assets are stored by third parties. A third party, whether it is a website, a company or a government, keeps silos of identity data. They often require more information than they need in order to verify individual's identity such as ``mother's maiden name'', ``phone number'' and ``social security number.'' Sensitive identity details are often stored in repositories that identity owners are unaware of, shared without their approvals, and exploited for commercial purposes.

\subsection{Security}
In traditional systems, each service provider keeps some portion of individual's identity information for identity verification. Hackers constantly attack these systems to steal the identity information. Potential data breaches result in tremendous setbacks for both the identity owners and the businesses. According to Javelin’s Identity Fraud Study \cite{Javelin2018} over $16.7$ million customers in the U.S. were affected from identity frauds, which cost them a total of $\$16.8$ billion in 2017 alone. 

What's more interesting is that the number of victims affected by identity frauds was increased by $8\%$ when compared to the previous year. There is an increasing trend in cyber-security breaches and a rise in their economic damages. A report by Herjavec Group predicts that damages caused by cyber-security breaches will cost the world \$6 trillion in a year by 2021 \cite{CyberCrimeReport_Herjavec}.

\subsection{Globalization}
From global perspective, identity verification and record attestation are challenging tasks across borders due to the institutional and international barriers. When a person travels to a different country, identity verification often starts from scratch and boils down to a manual process of verifying his/her physical documents (i.e., passports.) In addition, verification of credentials such as education certificates, diplomas and credit reports is often slow. Also, disconnected processes requires involvement of multiple third parties for attestation. For instance, individuals who have earned a particular college degree in India have to go through trusted third parties, costing them a significant amount of money and time in order to prove their degrees in a U.S. governmental office.

\section{Concepts and definitions}
\label{concepts}

\subsection{Blockchain}
\label{blockchain_concept}
Blockchain is an immutable distributed ledger that stores ownership of digital assets in the form of transactions and blocks. A Blockchain network consists of peers, each keeping the same copy of the ledger data managed through peer-to-peer (P2P) networking. Unlike traditional decentralized peer-to-peer networks, in which each peer acts independently, in Blockchain final decision is made through a consensus among the peers. Blockchain was first introduced as the backbone technology in Bitcoin, a digital currency system which eliminates the necessity of trusted third parties in electronic payments but still guarantees trust among the peers \cite{nakamoto2008bitcoin}.

In a Blockchain ledger, transactions are ownership transfer of digital assets. 
In Bitcoin system, ownership transfer is called ``payment,'' in which digital asset is the digital currency (also called Bitcoin), and the owners are Bitcoin wallet holders identified by asymmetric cryptographic keys. Specific peers called miners are responsible for approving transactions. When a transaction is initiated, transaction details including sender, receiver, digital asset amount being transferred, and the timestamp indicating the time of the transaction are hashed and broadcasted to the pending transaction pool. Miners grab a list of transactions to constitute a candidate block. The candidate block also contains a timestamp, the hash value of the candidate block which is generated from the block header and the Merkle root hash value \cite{merkle1980protocols} of transactions included in the block, and the hash value of the last approved block in the ledger. 

%Serkan: cited PoW paper
The Blockchain network utilizes a consensus algorithm to approve and append candidate blocks to the ledger. For instance, in Bitcoin network the proof of work (PoW) consensus algorithm \cite{vukolic2015quest} is used, in which miners compete to approve candidate blocks by trying to compute a computationally hard Nonce value. The Nonce value is determined by the cryptographic consensus algorithm and an ever growing difficulty of the network. Once a block is validated, it is appended to the previously approved blocks by referencing to the last block's hash value, constituting a chain of approved blocks in the ledger. 

The chained and distributed mechanism of Blockchain makes tampering the previously approved transactions impractical. Because tampering a single transaction in a block would result in a different Merkle root hash value of the transactions contained in the block, 
and a different hash value for the tampered block. Therefore, it invalidates all the following blocks in the chain due to the hash linking feature of the ledger. This requires the re-calculation of a different Nonce value for all the blocks including and after the tampered block in the chain for the current peer. In order to verify the fake transaction, aforementioned process must also be done for the majority of the peers in the network. It is computationally impractical in a widely adapted Blockchain network. Thereby, the ledger data stored in a Blockchain network is typically considered immutable.

\subsection{Cryptography, hashing and digital signature}
\label{pki}
Public-key cryptography (asymmetric cryptography) \cite{rivest1978method} is an encryption technique that has been used for decades. It makes use of a key pair consisting a private and a corresponding public key. Asymmetric cryptography enables encryption and decryption of messages using two separate keys, in a way that a message encrypted with a public key can only be decrypted with the relevant private key that belongs to the key pair, and vice-versa. While private keys are meant to be only known by the key owners, public keys are open to others. In Blockchain, public-key cryptography is utilized for asset ownership and for verifying the authenticity of transactions.

%Serkan: background kismi cok uzun. kisaltmak icin bu figure ve paragraph comment out edildi.
%Figure \ref{fig:fig-public-key-cryptography1} shows an example of public-key cryptography utilization in message sending for encryption and decryption. In the example, sender Bob uses receiver Alice's public key for the purpose of encrypting a message, and delivers the encrypted message to Alice. The encrypted message can only be decrypted by means of Alice' private key. The mechanism ensures that only Alice is able to retrieve the original message as long as Alice maintains her private key.

%\begin{figure}
%	\includegraphics[scale=1]{fig-public-key-cryptography1}
%	\caption{Mechanism of public-key cryptography.}
%	\label{fig:fig-public-key-cryptography1}
%\end{figure}

Hashing is a mathematical mechanism of generating a fixed size value from input data. It is impractical to regenerate the original input given its hash value. In Blockchain, hashing is utilized in hashing transaction and block data. For example, Bitcoin system uses SHA256 hashing algorithm which was originally designed by United States National Security Agency (NSA). SHA256 is also being used for decades in many sectors and services that require solid security such as in financial services, and it has proven to be secure. Table \ref{tab:sha256hash} shows sample inputs and output of SHA256 hash function.

\begin{table}[h!]
	\caption{Sample inputs and outputs of SHA256 hash function.}
	\begin{center}
	  {\renewcommand{\arraystretch}{2.2} 
	  \begin{tabular}{|l|l|}
			\hline
			\textbf{Original text}& 
			\textbf{SHA256 hash}
			\\
			\hline
			``tubitak''& 
			\pbox{8cm}{8c9b3371a4cae382bad1d752000902f871f8f78b\\1a2b62e4fe3ac47f40a2b742} \\			
			\hline
			``Tubitak''& 
			\pbox{8cm}{50ae8005208300584bd519ecfca19a083ad2831\\930668cee1b594bc8bb1b353c} \\			
			\hline
			\pbox{5cm}{``The Scientific and Technological Research Council of Turkey (TUBITAK)''}& 
			\pbox{8cm}{e18dd11e01d89410631d22829ea7786c6422878\\5669d6a5f665e33fa348f3fc2}\\
			\hline
		\end{tabular}
		}
		\label{tab:sha256hash}
	\end{center}
\end{table}

\begin{figure}
	\includegraphics[scale=0.95]{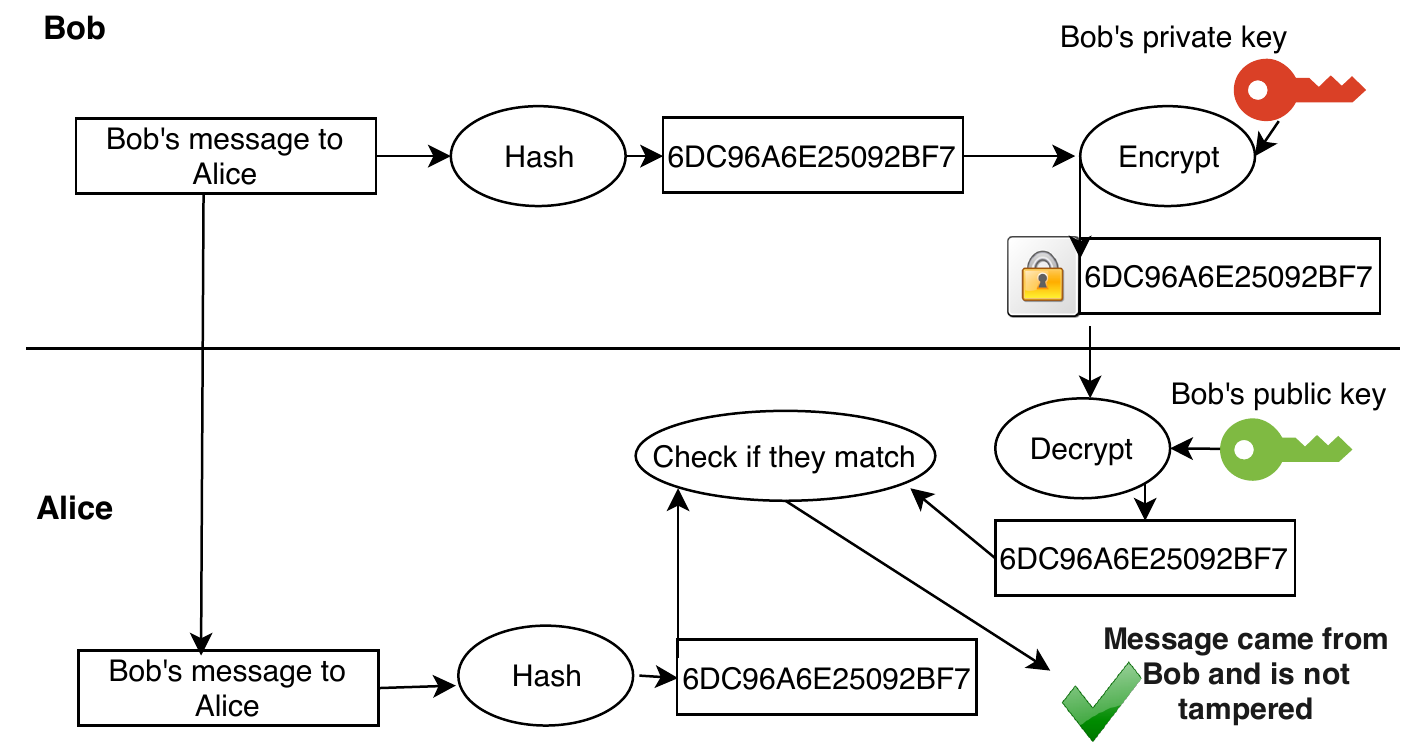}
	\caption{Sending data using digital signature.}
	\label{fig:fig-digital-signature1}
\end{figure}

Digital signature refers to digitally signing a message in order to certify authenticity and ensure integrity of the message in peer-to-peer communication, which is achieved using public-key cryptography and hashing. As an example, in Figure \ref{fig:fig-digital-signature1}, Bob sends a message to Alice using digital signatures. Bob encrypts the hash of original message with his private key. This action is called signing. Bob sends original message and signed message to Alice. Alice applies the same hash function to the original message and decrypts signed message using Bob's public key. Alice then compares these two results. If they are equal, then it means that the message was indeed came from Bob and was not tampered. In Blockchain, digital signatures are utilized in verifying authenticity of digitally signed transactions. For instance, in Bitcoin system, only transactions initiated by the asset owner itself are verified, which means that asset owners can only spend the assets they own. This is ensured via the concept of digital signatures.

%\subsection{Identity Standardization}
%\label{id-standardization}

\subsection{Self-sovereign identity}
\label{ssi}
Self-sovereign identity (SSI) is an identity model, in which any person, organization, or entity has the ownership and full control of their own data. It is not governed by centralized authorities, and it can never be removed from the identity owner. The requirements of an SSI model are described as below \cite{abraham2017self}:

\begin{itemize}
	\item Identity owners have full control over the data they own.
	\item Integrity, security and privacy of owner's identity are ensured by the system, a central authority is not required for trust.
	\item Provides full portability of the data. This means that identity owners can use their identity data in where they want (for instance in accessing an online service.)
	\item Any changes to the data is transparent, and transparency is sustained by the system.
\end{itemize}

\subsection{Decentralized identifier}
\label{did}
Decentralized identifier (DID) is an identification mechanism which assigns a standard, cryptographically verifiable, globally unique and permanent identity to individuals, organizations, and things. DIDs are completely under the identity owner's control and do not depend on central authorities. Public-key cryptography is used in DID as each DID contains an asymmetric key pair of a public and an associated private key. The control of a DID is managed through the DID's private key. DIDs provide an identity owner a lifetime long encrypted private channel with another identity owner. Identity owners use DIDs to identify themselves. Each DID resolves to a DID document (DID descriptor object), which contains DID's cryptographic keys, publicly available metadata (if any) regarding the DID owner, and resource pointers for the discovery of endpoints for initiating interactions with the DID owner.

\subsection{Verifiable credentials}
\label{VerifiableClaims}
Credentials are proof for identity owners to assert their license or qualification on certain subjects. They are widely used in individuals' daily lives. Driver's licenses, university diplomas and travel passports are examples of the credentials. Verifiable credentials are machine readable, privacy respecting, cryptographically secure digital credentials of identity owners. Verifiable credentials support self-sovereign identity, such that identity owners accumulate credentials into an identity account and use the credentials to prove who they are.

Verifiable credentials usually involve a third-party attestation, but they can also be self-attested. Attestation is done by utilizing the concept of digital signatures. An attester (issuer) having a DID creates a verifiable credential by signing identity owner's records using its private key. Then, the credential is cryptographically verified by a verifier using the attester's public key. 
Verifiers rely on the credibility of issuers when issuing the trust on the credentials.

%Serkan: reviewer tavsiyesine göre comment ettim. Bununla ilgili çalışmada çok detay yok 
%\subsection{Zero-knowledge proof}
%\label{zeroknowledgeproof}
%Goldwasser et al. (1985) \cite{goldwasser1989knowledge} introduced zero-knowledge protocol as a technique by which a prover is able to evidence to a verifier that information being present is true, with hiding details of the information except from the fact that revealed information is in fact true. The method involves interactions between the prover and verifier with probabilistic calculations. In Bitcoin system and in regular Blockchain implementations, transactions details including the sender, recipient and the amount of the digital asset being sent are visible to all the participant in the network. Modern Blockchain implementations such as Zcash \cite{hopwood2016zcash} created privacy-preserving transaction systems by integrating zero-knowledge protocol, which hides transaction details and respect the privacy of the actors involved in the transactions.

\section{Proposed solution}
\label{proposed_solution}
%Maydar: burada corona ile ilgili, remote identity, credential verification and sharing olaylarindan bahset. kisaca.
We propose a Blockchain based digital identity solution, which makes use of attribute-based data sharing, self-sovereign identity, decentralized identifiers, verifiable credentials, and allowing identity owners to use trust relationships that they already have with trusted partners. 
The system enables identity owners to prove that they are who they claim to be (authentication, i.e., login systems) 
%, and making an explicit approval for an action, system, or transaction (authorization, i.e., wire transfers), 
and making a certain claim involving third-party certifications (attestation, i.e., proof of education degree certificates). 
The aim of the proposed system is to provide a framework that is remote-friendly, scalable, globally usable and providing both privacy and security by design.
%maydar, added remote-friendly here

%Genel olarak, 
%kişilere ve kurumlara Blockchain üzerinden dijital kimlik atandığı, 
%Kimlik sahiplerinin kendi kimlik ve belgeleri üzerinde daha çok kontrole sahip olduğu,
%Kişilerin kimlik ve kimliğe dayalı belgelerini ilgili kurumlardan aldığı ve mobil cihazlarında tuttuğu, 
%Ve ilgili kişi ve kurumların sunulan belgelerin originitesini onaylayabildiği bir sistem ön-görüyoruz.
%Veri koruma regulasyonları ile uyumlu, 
%kimlik sahiplerinin kendi bilgilerine bütünüyle sahip olduğu, 
%kimlik sahipleri hakkındaki bilgilerin standart bir şekilde tanımlanabildiği, 
%güvenli veri alışverişinin yapılabildiği ve standard bir şekilde doğrulanabildiği, 
%ve daha otomatize edilmiş bir iş akışına olanak sağlayan güvenli bir dijital kimlik modeli geliştirmektir
%Bu çalışmada, kimlik paylaşımını, kimlik doğrulamayı ve kimliğe dayalı veri paylaşımını geleneksel sistemlere göre daha etkili, 
%daha mahremiyetli ve daha güvenilir hale getiren Blokzincir tabanlı kimlik yönetimi ve tutanak tasdik sistemi öneriyoruz. 

\begin{figure}
	\includegraphics[scale=0.3]{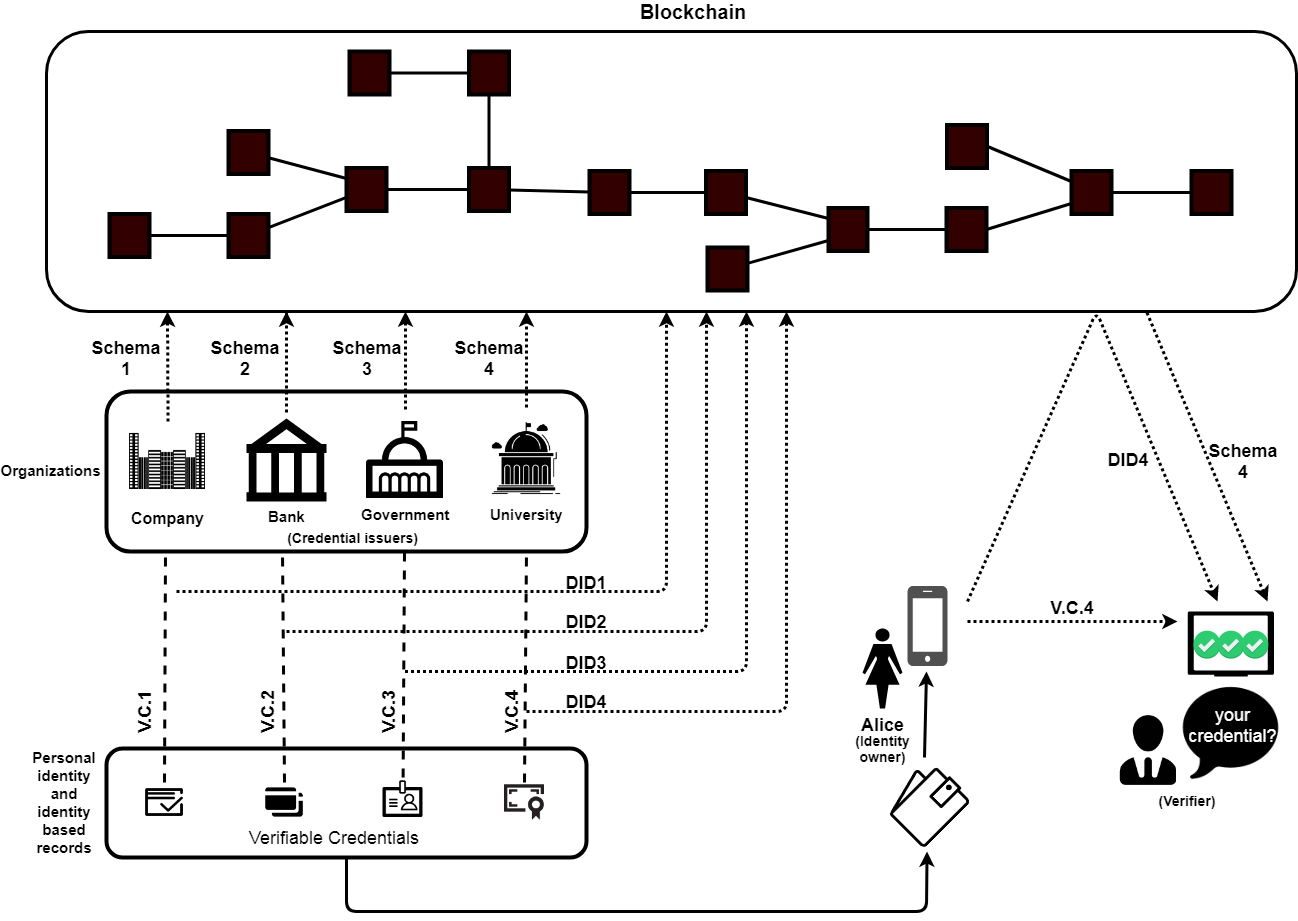}
	\caption{Overall workflow of the proposed identity system.}
	\label{fig:fig-overall-structure}
\end{figure}

Figure \ref{fig:fig-overall-structure} shows the overall workflow of the proposed solution. In the system, individuals and institutions are assigned to a digital identity through a decentralized platform orchestrated by Blockchain. The system does not store any private user identity data, not even encrypted version, in public ledger. 
It only contains the proofs of transactions for the verification in Blockchain.
Identity owners fully control the user identity data and solely determine whom to share their identity data. Identity owners receive their cryptographically signed documents in the form of cryptographically verifiable credentials from the related institutions, and keep them in their mobile wallets. 
Consequently, the system also reduces the time spent in tasks requiring verification of user identity and eliminates the need for central authority in verification and  management of identity data. 

%Serkan todo: itemize the contributions
The main benefits of proposed solution include eliminating the need for central authority for identity verification and identity data management, reducing the time spent in verification of identity, %preventing user password requirement for all services,
allowing data sharing with permission, and verifying origin of the data while sharing.

\subsection{Sample Use Cases}
\begin{figure}
	\includegraphics[scale=0.4]{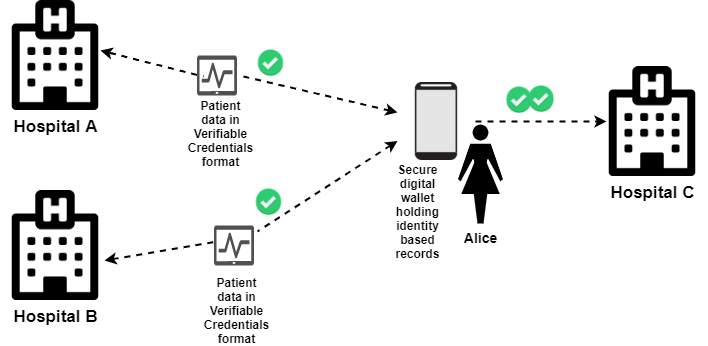}
	\caption{An example use case for patients medical records.}
	\label{fig-usecase-hospital}
\end{figure}

%Maydar: burada 3 tane use casi resimlerle birlikte include edip acikladim
In this section, we describe the proposed framework with sample use cases. The framework enables organizations to issue digitally signed documents to identity owners. The signed documents are in the form of verifiable credentials. Verifying parties are able to verify that the documents are original, not mutated and signed by the issuers with the help of digital signatures. Figure \ref{fig-usecase-hospital} shows an example use case regarding a patient's medical records involving multiple medical institutions. In the use case, a patient named Alice was previously treated in hospitals A and B, and presents her previous medical records from these institutions at hospital C. Hospitals A and B issue Alice's medical records to her in the form of verifiable credentials. Alice gathers these documents in her secure digital wallet, and presents to her physician at hospital C. Knowing hospital A and B's public keys, hospital C is able to verify that Alice medical records are indeed originally issued by hospital A and hospital B and are not mutated.  

\begin{figure}
	\includegraphics[scale=0.4]{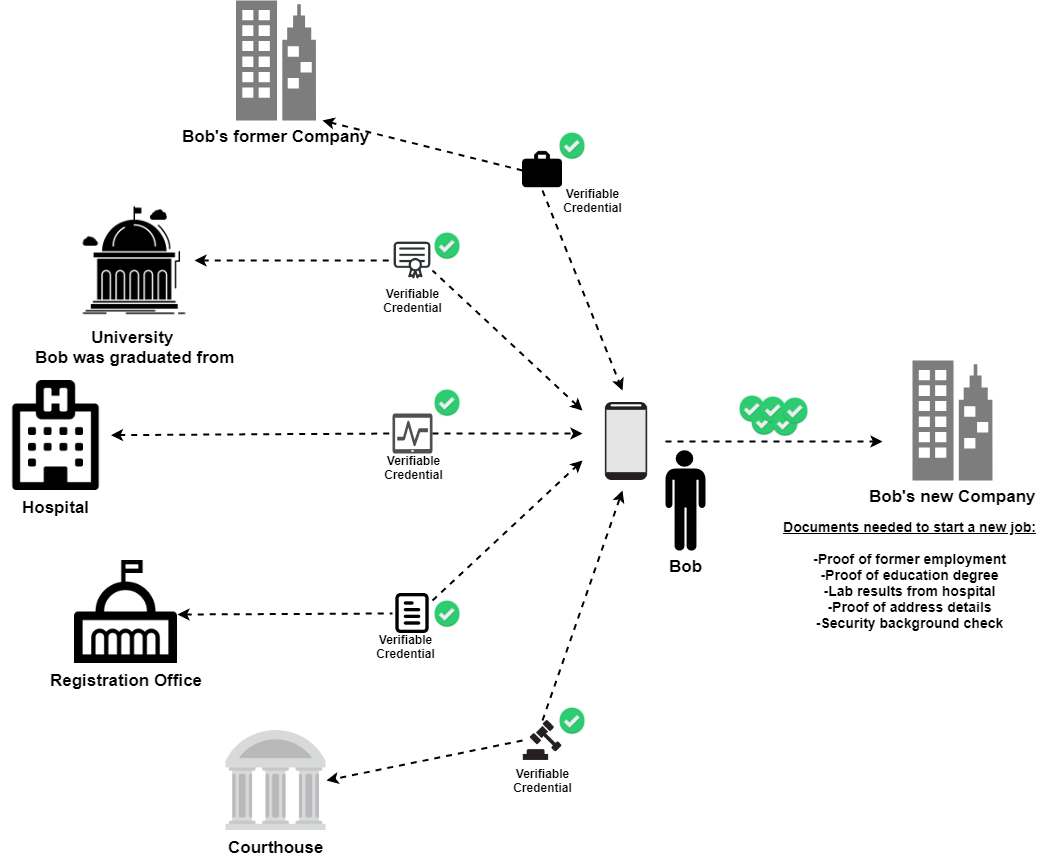}
	\caption{An example use case for presenting documents to a new employer.}
	\label{fig-usecase-start-job}
\end{figure}

Figure \ref{fig-usecase-start-job} shows an example use case for presenting documents and credentials to a new employer in order to start a new job. In the example, Bob's new company requires Bob to present proof of educational degree, proof of former employment, lab results from hospital, proof of address details and documents regarding background check. Bob gathers the documents from related institutions online in the form of verifiable credentials and keeps them in his digital wallet, and Bob's new company verifies the authenticity of the documents once presented. The process saves time, enables issuing and verifying documents online and prevents counterfeiting of records.
%maydar: yukarida  "enables remote works" demisim. daha uygun bir cumle bulamadim.
%maydar: burada, Bob yabanci bir ulkeden mezun oldu. fakat yeni ulkesindede dogrulanabilir referanslar yardimiyla belgelerini dogrulatabiliyor diyelimmi?

\begin{figure}
	\includegraphics[scale=0.5]{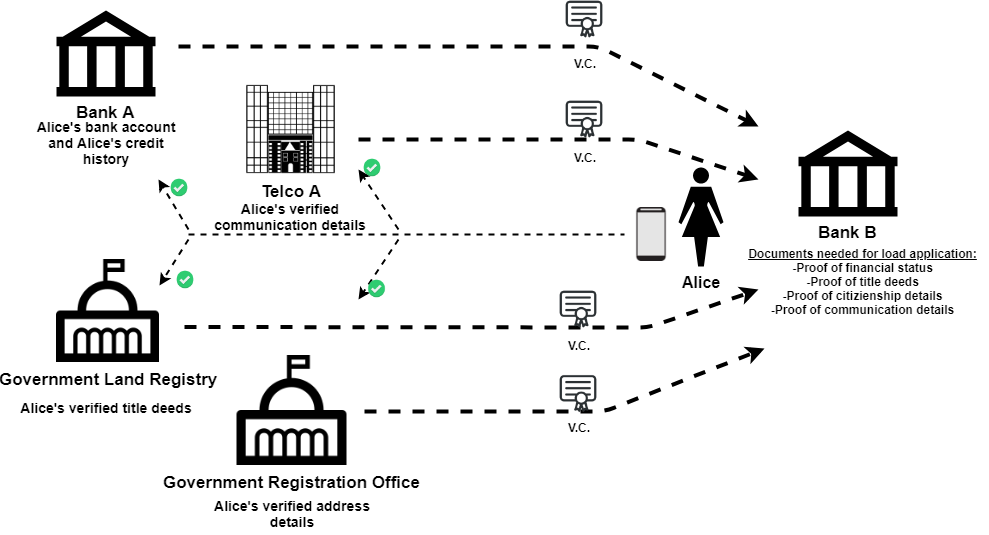}
	\caption{An example use case for a loan application.}
	\label{fig-usecase-loan}
\end{figure}

Figure \ref{fig-usecase-loan} shows an example use case regarding a loan application. In the use case, a loan applicant Alice applies for a loan with Bank B. Alice is a current customer of Bank A, which is trusted by Bank B. In addition, Alice owns a land and regarding information are kept by government departments. However, Alice has never worked with Bank B, previously. Therefore, Bank B does not have any records about Alice. As part of KYC regulations, Bank B is required to know Alice in order to serve her. Using the proposed framework, Alice authenticates herself with Bank B. Then, Bank B requests Alice's information from related organizations online, and she receives a notification regarding the request. Alice gives her consent for her information to be shared with Bank B in the form of verifiable credentials. Bank B processes Alice loan application based on the gathered information. The whole process happens online in minutes without Alice physically having go to the bank. 

%\begin{figure}
%	\includegraphics[scale=0.4]{fig-usecase-tapu.png}
%	\caption{An example use case regarding land registration.}
%	\label{fig-usecase-tapu}
%\end{figure}
%Maydar: tapu use casini eklemedim

%Serkan: Bu kismi Method bolumune tasiyip genislettim 
\subsection{Blockchain Network}
\label{blockchain_for_identity_management}
The backbone of our identity management system is the Blockchain network. 
%Blockchain technology is considered as a key innovation in the age of digitalization, in which all components are either digital or have a digital representation, and are connected. 
In the system, Blockchain network is used for bringing transparency and trust to digital ecosystem by assigning a digital identity, distributing the storage rather than centralizing, and automating the processes with smart contracts. The concept of ``decentralized digital identity'' is in line with the fundamental design of Blockchain.
%In Blockchain, asset owners are identified by asymmetric cryptographic (public-key cryptography) keys. Blockchain-based identity solutions utilize the concept of asymmetric cryptography for assigning digital identity to data representations. 
Namely, the following aspects of Blockchain technology made it the ideal choice for the proposed identity management system:

\begin{itemize}
	\item Blockchain ledger is immutable and transparent (based on permissions), which are essential parts of identity management. 
	\item Blockchain is resistant to single point of failure and denial-of-service attacks.
	\item Blockchain provides an efficient implementation of public-key cryptography and hashing, which: 
	\begin{itemize}
		\item can be extended for digital identity ownership.	
		\item helps ensuring integrity and authenticity of identity-based records.
		\item can be utilized for third-party attestation of records.
		\item helps facilitating permission-based record sharing with smart contracts.		 
	\end{itemize}
	\item Blockchain eliminates or diminishes monopoly in identity management, as it is not controlled by any central authority. This also enables identity and record integration in global scale.
	\item Blockchain supports incentives via crypto-currencies, which can be utilized for certain tasks such as providing incentives to the participants for data sharing.
\end{itemize}

%Serkan: Bu kismi blockchain network sectioni altina tasidim
\subsubsection{What is stored on Blockchain?}
\label{What is stored on Blockchain}
%Serkan: updated the section
For a Blockchain network, storage is a vital issue to be considered from perspectives of identity management, scalability, security and privacy. To avoid potential security and privacy problems, no private data is stored on the Blockchain ledger in our system. Even encrypted and hashed versions of private data are not stored as the encrypted data on Blockchain might become vulnerable to advanced quantum machines in the future \cite{tessler2017bitcoin}. Keeping sensitive private data on the ledger carries a risk that if the private keys of the identity owner are compromised, the identity owner's data can be revealed to public. Thereby, in our system Blockchain is mainly utilized for searching decentralized identifiers and identity owners. It only stores the consent proof of data sharing between the identity owners and the revocation registry. Since the proofs of data are stored on the Blockchain rather than the identity data themselves, the scalability is not an operational challenge.

%Serkan: Consensus mechanism sectioni eklendi
\subsubsection{Consensus mechanism}

The Blockchain network in our system utilizes Plenum Byzantine Fault Tolerance  \cite{hyperledgerindy-plenum_2016} consensus mechanism that is implemented by Hyperledger Indy. 
Plenum is developed based on the Redundant Byzantine Fault Tolerance (RBFT) algorithm \cite{aublin2013rbft}. The main idea of RBFT is that it enables running multiple instances of the Byzantine Fault Tolerance (BFT) \cite{lamport2019byzantine} protocol on different machines concurrently. One of the instances is promoted to be the master node, which has the authority to execute orders. The other instance(s) in the system maintain a replica of the ledger and can order requests. However, the updates to the ledger can only be executed by the master node. All backup instances track and compare their performances against the master instance. If the performance of master instance with regards to latency and throughput reduces below an acceptable threshold, the master is replaced by another backup instance \cite{aublin2013rbft}.  
 
Compared to proof-of-work (PoW) \cite{vukolic2015quest}, the RBFT based consensus mechanism performs better in terms throughput and speed. Due to nature of BFT algorithms, the time to reach consensus in RBFT increases with the size of the nodes in network. 
Although our consensus mechanism is not as scalable as Pow, its scalability is sufficient for a permissioned Blockchain network. 
The only major drawback of RBFT consensus method is the requirement that all nodes in the network must be connected and known by all other nodes. This introduces potential centrality to the network as the identities to the members of the network must be provided by a trusted party \cite{vukolic2015quest}. However, this is not a disadvantage in our case as the proposed system is based on a permissioned Blockchain network, in which the participants of the identity management system are known to the identity issuers.
 
\subsection{Digital identity management}
In the system, individuals and organizations identify themselves with self-sovereign identities, which they fully control their identity based records without relying on a central authority. It can also be extended to devices in internet of things (IoT) domain for providing device identification and authentication. SSI consists of multiple decentralized identifiers (DID), one for each relation the identity owner has with other identity owners. The advantage of using different DIDs for each relation is that in case the keys from a particular DID are compromised, the other DIDs of the user stay protected. Under identity owner's control, each DID is globally unique and includes a cryptographically verifiable PKI (public, private key pair). Each DID resolves to a DID document, a DID descriptor object (DDO) which is stored on the Blockchain. A DDO includes the public key associated with the corresponding DID and metadata needed to prove ownership the corresponding DID, and endpoints of the DID objects to initiate trusted peer interactions between the ledger entities. 

\begin{figure}
	\includegraphics[scale=0.71]{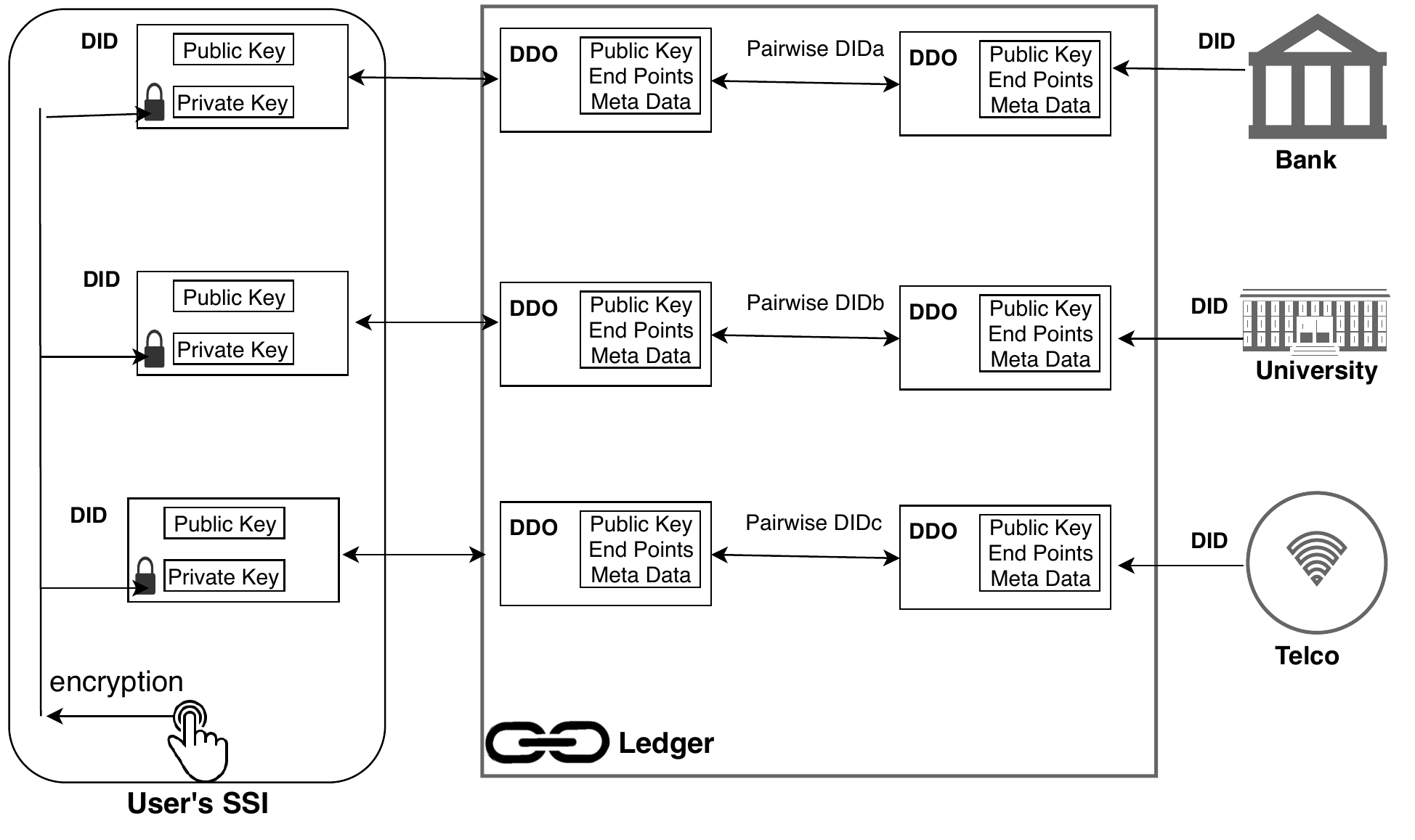}
	\caption{A demonstration of how self-sovereign identity is stored on user device and pairwise decentralized identifiers.}
	\label{fig-sol-new-user-did-ddo}
\end{figure}

The details of cryptographic key pairs (public and encrypted private keys) of DIDs that belong to the user's self-sovereign identity are stored on user's devices, such as mobile phones. While public keys are stored in non-encrypted form, corresponding private keys are stored in encrypted form. A private key is encrypted in a way such that it can only be decrypted using a biometric signature of the identity owner, as fingerprint, facial feature, an iris or a retina. Encrypting private keys provide multi-factor and identity owner-specific authentication in order to be allowed access to the identity details. Figure \ref{fig-sol-new-user-did-ddo} shows an example of how self-sovereign identity is stored on a user device and pairwise decentralized identifiers in the system.

\subsection{Authentication mechanism of the system}
\label{authentication}
Authentication is the process for identity owners to prove ownership of their identity. It is often required in individuals' daily lives for purposes such as security checks, granted access to specific services. Our system uses public-key cryptography based authentication.

%\subsubsection{Traditional authentication}
%\label{authentication-regular}
%In our system, a user can use a regular username and password based authentication mechanism, if the user already has an account with the verifier. In this case, the DID of the user is linked to the user's account, and the verifier can still locate the DDO record of the user and can use the user's public key for additional tasks such as record sharing. 

\begin{figure}
	\includegraphics[scale=0.87]{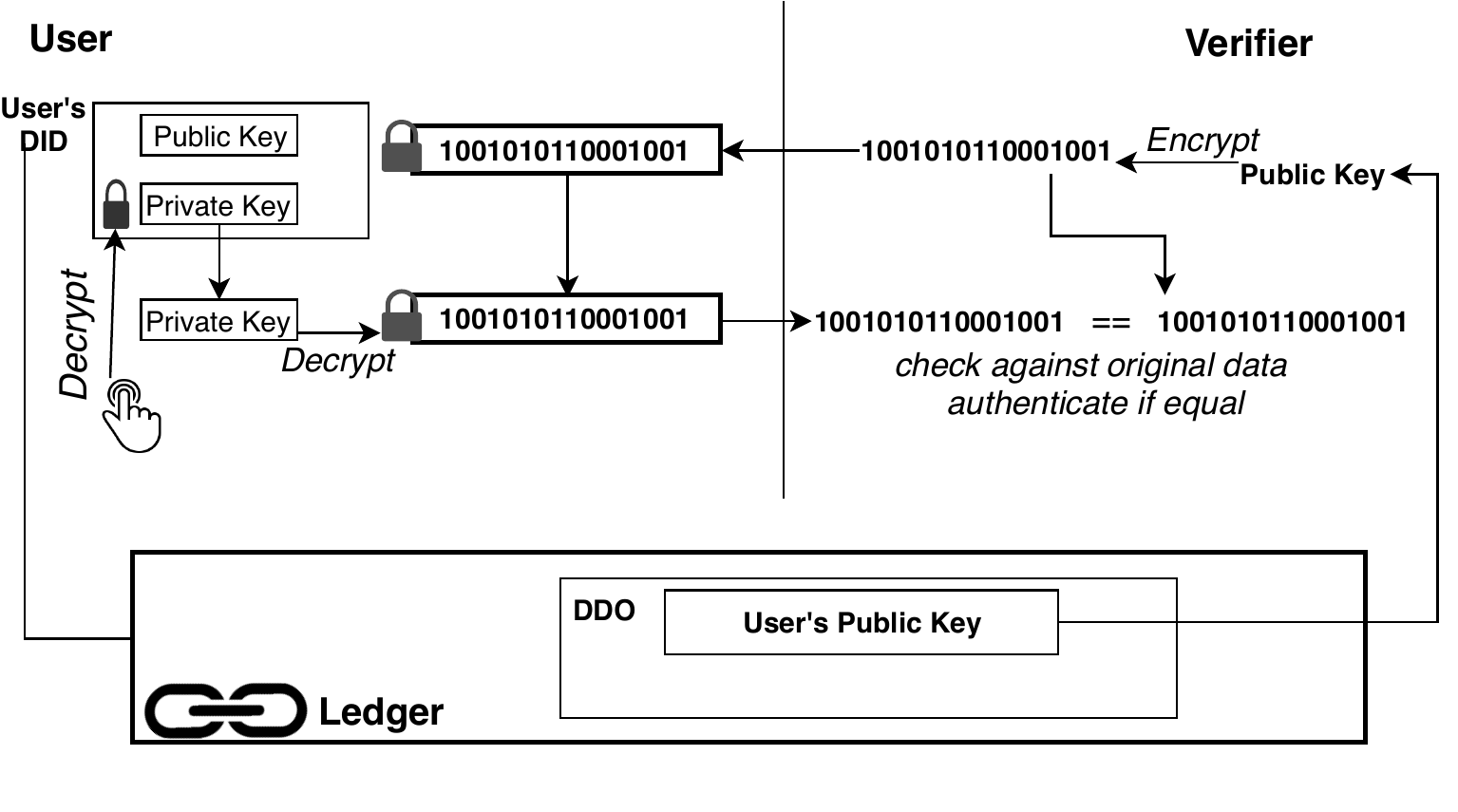}
	\caption{Authentication using public-key cryptography.}
	\label{fig-sol-authentication}
\end{figure}

%\subsubsection{Public-key cryptography based authentication}
%\label{authentication-pki}
Identity owners are required to prove that they have the control of the private key of a public key associated with their identity. Verifier side (for instance a website) encrypts a random string with this public key and sends it to identity owner. Using the private key, identity owner decrypts and retrieves the original string. Then, it sends the string back to the verifier. The verifier checks it against the original string, and authenticates identity owner if they are equal. Figure \ref{fig-sol-authentication} illustrates how public-key cryptography based authentication works in the system.

\subsection{Verifiable credentials in the system}
\label{verifiable-claims}
In proposed system, identity based record sharing is achieved through the concept of verifiable credentials. 
%As explained in section \ref{VerifiableClaims},
Credentials contain data about an identity owner regarding their license or qualification on certain things. Each credential needs to be digitally signed by the issuer of credential. There could be two type of credentials; a self-attested credential is issued and signed by the identity owner itself (i.e., user publishes and signs his/her own data), and a third-party credential which is generally issued and signed by a third party (i.e., attestation and notarization services.)

Credentials data can be in the form of free-text, graphic or pre-defined credential definition (schema). Our system encourages identity owners and third-party credential issuers to use a pre-defined credential definition in order to make the content of credentials machine readable. Data schemas are important for defining and making credentials machine readable. Our system allows schemas to be published on the ledger. By making use of RDF and ontologies, a great deal of already defined schemas can be utilized such as ontologies regarding personal data, medical data and university diplomas. 

For a credential to be issued by third parties (issuers), issuers first need to authenticate identity owners. Once authenticated, issuer picks an appropriate credential definition, constructs and signs the record with its private key, and delivers the signed record to identity owner.
As an example, by using our system, driver's licenses can be issued digitally in the form of verifiable credentials. For this, a government department needs to publish a credential definition for driving licenses onto the ledger. The credential definition contains references to attribute names and types from credential schema, which holds information such as the driver name, license number, license issue and expiration dates and license type. A license authority receives the license schema from the ledger, fills out driver's information accordingly, and cryptographically signs the form, which generates a digital version of a driving license in the form of verifiable credentials. As a result, the license owner keeps the digital credential on his/her devices. Authorities are able to verify that the driving license is owned by the driver, was signed by a legitimate license authority and is valid. Credential definitions stored in the ledger are indexed and made discoverable. This system enables users to identify the authorities or organizations, which issue credential definitions.

\begin{figure}
	\includegraphics[scale=0.56]{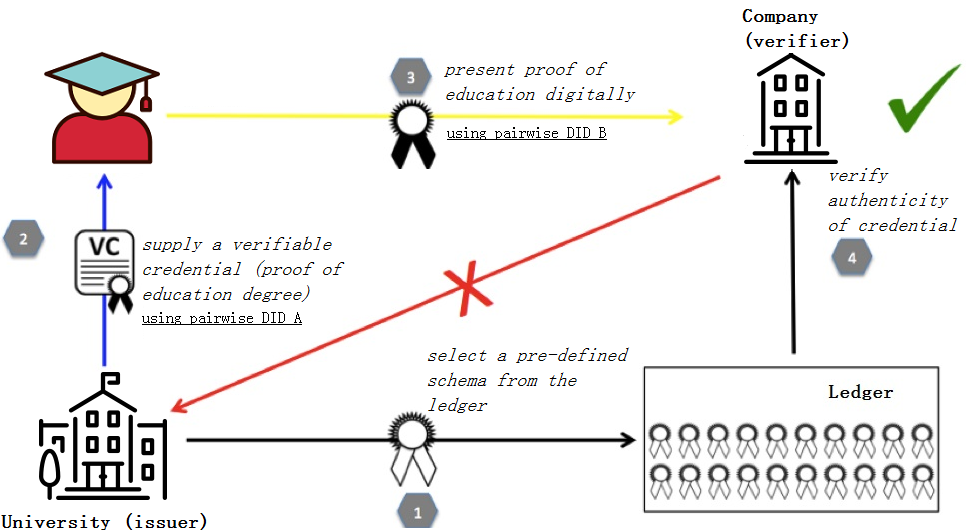}
	\caption{An example of a verifiable claim interaction.}
	\label{fig-sol-verifiable-claim-example}
\end{figure}

Verifiable credentials can be exchanged digitally between identity owners, involving individuals and institutions. Figure \ref{fig-sol-verifiable-claim-example} shows an example of exchanging credentials of a University degree certificate. In the example, a University supplies its former student a verifiable credential proving her educational degree, using an existing claim schema from the ledger, and via pairwise decentralized Identifier ``A''. The University graduate presents the digital credential to her company using pairwise decentralized identifier ``B''. The company confirms that her employee's educational credentials are valid by verifying authenticity of the verifiable credential.

A verifiable credential can also be issued upon a request from another third party. In this case, identity owners have already proven their identity with an institution (credential provider), and also need to prove their identity with another institution (credential requester,) using the trust relationship they have with the credential provider. 
In order to do so, credential requester authenticates identity owner, and redirects identity owner to the authentication system of credential provider. Based on the requested information, credential provider provides a verifiable credential to identity owner. 
Identity owner signs the verifiable credential with his/her private key, and forwards it to the authentication system of credential requester. 
Credential requester retrieves the DID of identity owner and DID of the credential provider, and verifies digital signatures of them using their public keys from DIDs. 
This whole process can be performed online in seconds, enabling identity owners digitally proving their identity and credentials to credential requesters. 

\begin{figure}
	\includegraphics[scale=0.8]{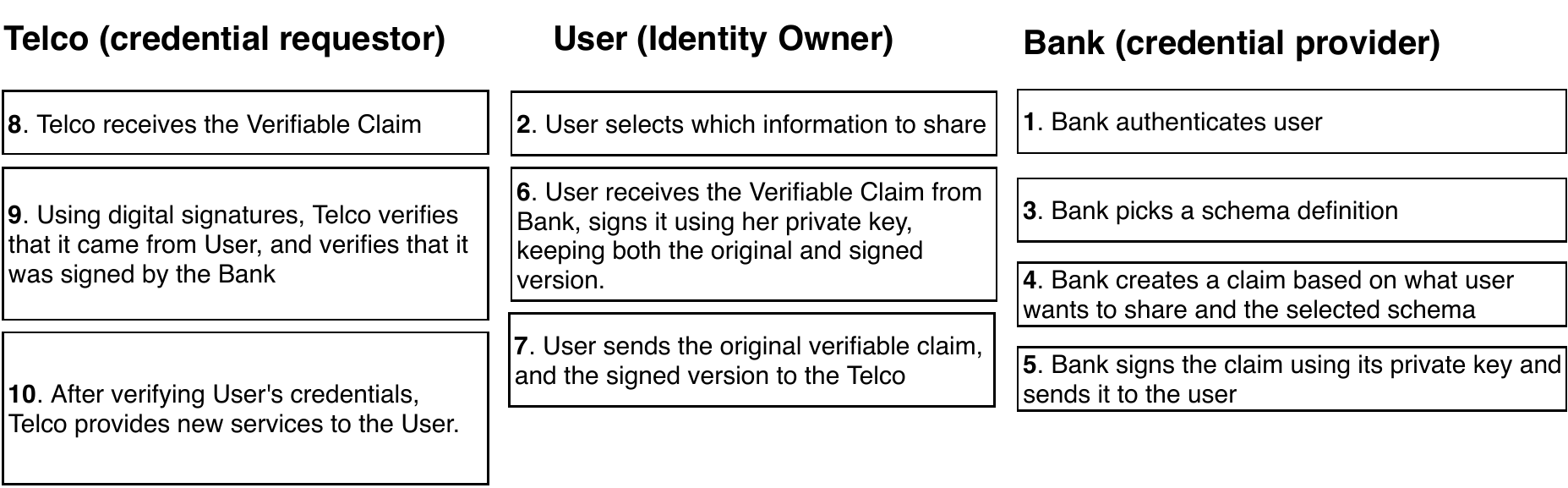}
	\caption{Using Verifiable Claims for third-party attestation.}
	\label{fig-sol-verifiable-claim1}
\end{figure}

The system also assists in satisfying Know Your Customer (KYC) requirements by enabling organizations linking their online services with verifiable credentials, while ensuring that information is only shared with the consent of identity owners. 
A consent proof of these actions is stored on the ledger without revealing sensitive information of parties involved.
For instance, if identity owners have proven their identity with their bank, they can grant permission to share their financial data such as credit score with a telecommunication company in order to request a new service from the telecommunication company as illustrated in Figure \ref{fig-sol-verifiable-claim1}.

\subsubsection{Revocation registry}
Verifiable credentials are issued to and kept by identity owners. Verifying a credential involves confirming that the credential is owned by the identity owner and issued by a trusted authority, and that the credential is still valid, i.e. it has not been revoked by the issuer. Therefore, an efficient revocation mechanism is needed which does not put a lag on the system (asynchronous), respects the privacy of the identity owner (private) and is not controlled by central authorities (decentralized.)

Our system keeps a revocation registry on the ledger. Credential issuers are responsible to publish revoked credentials to the revocation registry. The registry is a cryptographic accumulator, which includes credentials or the specific attributes of a credential that have been revoked, in addition to the corresponding credential definitions. A cryptographic hash value of a revoked credential is calculated and kept in the revocation registry. Verifiers check the existence of the hash value to test whether the credential has been revoked or not.

\subsubsection{Consent manager}
When an identity owner shares a verifiable credential with a verifier, a proof of the sharing agreement (consent receipt) is generated and kept on the ledger. A concept receipt is signed by both identity owner and the verifier, and it includes their DIDs and the shared attributes names and data types without any of private information. The proof of a concept receipt is a cryptographic hash of the receipt, which is stored on the ledger. By this mechanism, tamper-proof evidences of sharing agreement of verifiable credentials are maintained, in case they are required in the future.

\section{Related work}
\label{related_work}

%Serkan: Bu uc paragrafi related work kismina tasidi
%\subsubsection{Challenges of Blockchain in identity management}
%\label{challenges_blockchain_for_identity_management}
It is noteworthy to state that in order to fully exploit the functionalities of a Blockchain-based digital identity solution, a mainstream adoption of the system is crucial. Although Blockchain technology enables novel innovations in the area, Blockchain has not yet reached a large scale adoption in its current technological state. 
According to a survey by Johansel et al. \cite{johansen2017comprehensive}, the scalability bottlenecks, the lack of the progress in accessible data and APIs, and the issues with disk space and bandwidth are the major hurdles of Blockchain for a conventional adoption. Swan \cite{swan2015blockchain}, on the other hand, presented technical limitations for adaptation of Blockchain in the context of security, usability, size and bandwidth, throughput and latency, versioning problems and wasted resources in transaction approval. Li-Huumo et al. \cite{yli2016current} revealed that Blockchain research studies mainly focus on Bitcoin system, and only $20\%$ of the studies concentrate on obstacles of Blockchain technology from security and privacy point of view. 

Zheng et al. \cite{zheng2016blockchain} brings up the concerns regarding the privacy of individuals in conventional Blockchain networks such as Bitcoin by stating that associating identity owner's pseudonym identifiers to IP addresses is possible \cite{biryukov2014deanonymisation}, and that transactional privacy cannot be guaranteed by conventional Blockchain networks \cite{meiklejohn2013fistful,kosba2016hawk}. 

Despite, there has been significant progress in Blockchain technologies and their applications to digital identities in recent years. Several reports \cite{JapanInstitute2016,blockchainUSpostal,walport2016distributed} originated by governments state the disruptive potential of Blockchain and the opportunities of exploiting the technology for digital identity.

%Serkan: Hyperledger kisimlari kisaltildi ve degistirildi
%\subsection{Hyperledger}
With open source community perspective, Hyperledger initiative, hosted by The Linux Foundation and members from diverse industries, led to the development of common open source Blockchain frameworks and tools that are publically available. As an umbrella project of the initiative, Hyperledger Fabric \cite{HyperledgerOverviewWhitePaper,cachin2016architecture} is a Blockchain foundation for creating private permissioned Blockchain applications with a modular design. %The members of the initiative are from a diverse industries including banking, technology, consulting, retail, governments and academia. 

%\subsubsection{Hyperledger Fabric} 
%\label{fabric}
A Hyperledger Fabric based Blockchain system includes actors such as client applications, asset owners, orderers and peers; each having a digital identity complies with X.509 digital certificate standard \cite{housley1998internet}. In Hyperledger Fabric, identities are essential since apart from asset ownership, management of resource and access permissions is also determined based on actor identities. Hyperledger Fabric also has the concept of principal, which includes additional properties of actor identities such as identity owner's unit, organization and permissions. %Verifiable digital identities in Hyperledger Fabric are provided through Certificate Authorities (CAs) within the network. 
%Serkan: Bu cumleyi comment out ettim. related work icin cok detay olmus.
%A Certificate Revocation List (CRL) is also kept by the system which stores digital identity certificates that have been revoked. Hyperledger Fabric also has a built-in Membership Service Provider (MSP) which identifies legitimate digital identities of the Blockchain network, and represents the organizations in which a digital identities belong. 
Another distinctive feature of Hyperledger Fabric is that it allows private channels that can be used for permissioned private data sharing.

%\subsubsection{Hyperledger Indy} 
%\label{indy}
Another Hyperledger project primarily built for self-sovereign decentralized identity supporting privacy by design, Hyperledger Indy \cite{HyperledgerOverviewWhitePaper} is a public-permissioned distributed ledger project. 
It develops a set of identity specifications, artifacts, libraries, tools, and reusable components for creating decentralized identity on Blockchain to enable identity interoperability across applications and distributed ledgers.
Hyperledger Indy supports data minimization. It enables identity owners to store their identity based records. Applications don't need to store individuals' personal data, instead they store a link to the identity. This way, identity owners are able to control access to their personal data. Hyperledger Indy's identity model supports decentralized identifiers and verifiable credentials. Our system was also developed based on Hyperledger Indy framework due to many built in overlapping features.

%\subsection{Sovrin}
\label{sovrin}
In a similar aspect, Sovrin \cite{SovrinWhitePaper} was offered as a live distributed ledger built for decentralized identity, which uses Hyperledger Indy's codebase. Sovrin ledger is a public resource that is designed to provide a self-sovereign digital identity for all. However, its governance model is permissioned. It means that the Blockchain nodes in Sovrin are governed by private organizations called ``stewards.''
Sovrin makes use of decentralized identifiers and verifiable credentials. To increase scalability, Sovrin uses two types of Blockchain nodes: A network of validator peers which have transaction write access to the ledger, and a bigger network of observer peers storing write-protected copy of ledger to handle requests for read.

According to Sovrin, decentralized identifiers and associated DID documents with verification keys and endpoints, schemas and credential definitions, proof of consent for data sharing, public credentials and revocation registries are stored on the ledger, whereas private data of any kind and private proof of existence are not stored on the ledger \cite{SovrinWhitePaper_whatgoesonledger}. The main difference between Sovrin and our solution is that Sovrin does not provide private key encryption and recovery mechanism for the private keys.

%\subsection{SecureKey}
%\label{securekey}
SecureKey \cite{SecureKey_bank,SecureKey_telco} is another Blockchain based identity and authentication provider similar to the proposed system. It allows customers to assert their identity information online using trusted providers that they have already completed the KYC process through trusted third parties such as government agencies, telecommunication companies and banks. SecureKey ensures that personal data is privately shared with explicit consent of identity owner. 
%Figure \ref{fig-secure-key-bank-user-experience1} illustrates an example of identity information sharing with customer consent, in which a consumer visits a telecommunication provider website, selects the device she desires, consents to share her identity information and her credit score provided by her bank, and the telecommunication company verifies her information in order to approve the requested service.

%Serkan: Related work kisminda figureleri cikarmak lazim. bu kismi cok uzatiyor, makalenin odagi kayiyor. related work kisa ve oz tutmak lazim.  
%\begin{figure}
%	\includegraphics[scale=0.52]{fig-secure-key-bank-user-experience1}
%	\caption{SecureKey: the user experience (Taken from \cite{SecureKey_telco}).}
%	\label{fig-secure-key-bank-user-experience1}
%\end{figure}

SecureKey uses a permissioned Blockchain network based on Hyperledger Fabric, in which participating organizations such as banks are central in managing the nodes of the Blockchain network. In the Blockchain ledger, the proof, provenance and permissions are stored. Consumer's identity information remains to be stored at the trusted providers. SecureKey also uses an incentive mechanism to data sharing, such that the credential requester pays to the credential provider. SecureKey architecture enables privacy with a triple-blind identity sharing, in that the data provider never knows the service a consumer is accessing, and the data requestor does not have to know the exact credential provider other than knowing the business type of the credential provider. However, SecureKey does not support verifiable credentials and self-sovereign digital identity for individuals. 

%\subsection{ShoCard}
%\label{securekey}
From identity owner empowerment perspective, ShoCard \cite{ShoCard} project was offered as a Blockchain based identity management platform, in which identity details are stored in a digital file called ``ShoCard''. The data is fully owned by identity owner and usually stored on the owner's mobile device. Identity owners have a public-private key pair for controlling of their identity. 
ShoCard system enables attribute based data sharing. The identity details are broken into multiple separate attributes. Each attribute is hashed and then signed by private key, and sent to be stored in a Blockchain network. Different than our system, private data is stored in Blockchain network in ShoCard, which might result in potential privacy implications. %Figure \ref{fig-shocard-attributes-hashing-certifications-stored-on-bc} illustrates ShoCard's attribute based self-certification system.

From a different aspect, Walmart filed a patent \cite{high2018obtaining} to protect a method that allows obtaining Electronic Health Record (EHR) of an individual from a Blockchain database even the individual is unable to communicate. In that system, personal medical data is managed on Blockchain. Individuals have access to their own record by controlling an asymmetric key pair, which is specific to their identity. The system is particularly useful in cases of emergency, in which the patient is unconscious or incapacitated and unable to provide the physician with critical information about pre-existing conditions or allergies that may influence treatment options.
% Figure \ref{fig-walmart-system-overview} illustrates the architecture of the system. 

%\begin{figure}
%	\includegraphics[scale=0.33]{fig-walmart-system-overview}
%	\caption{Architecture proposed in the Walmart's patent application for obtaining EHR stored in Blockchain (Taken from \cite{high2018obtaining}).}
%	\label{fig-walmart-system-overview}
%\end{figure}

In their system, the public key along with an encrypted private key based on a bodily feature of the patient, are stored in a wearable device. Patient's public key and the encrypted form of the associated private key can be obtained by scanning the wearable device via RFID. A separate biometric scanner device is used to obtain a bodily feature such as finger print or retina of the patient, and the encrypted private key can be decrypted using the biometric signature of the patient. %The asymmetric key pair enables accessing the medical records of the patient stored in the Blockchain. Figure \ref{fig-walmart-steps-to-access-patient-medical-record} shows the steps needed to access EHR stored in Blockchain.

%Serkan: commented out the following
%\begin{figure}
%	\includegraphics[scale=0.35]{fig-walmart-steps-to-access-patient-medical-record}
%	\caption{Steps to access EHR stored in Blockchain in the Walmart's patent application for obtaining EHR stored in Blockchain. (Taken from \cite{high2018obtaining})}
%	\label{fig-walmart-steps-to-access-patient-medical-record}
%\end{figure}

%\subsection{Bitnation}
In another related work, Bitnation \cite{jacobovitz2016blockchain} provides identity registration on Blockchain in order to enable geography-independent world citizenship unbound by governments. Bitnation can provide services like world citizenship, Blockchain passports, marriage certificates and emergency identifiers for refugees. 
A Bitnation identity requires concretely proving that the candidate existed at a definite time and location, and his/her existence was cryptographically signed by another group of identity owners. Bitnation uses Ethereum \cite{buterin2017ethereum} network, and utilizes hashing, digital signatures and smart contracts.

Bitnation partnered with Estonian government for Estonia e-residency program \cite{sullivan2017residency}, 
which offers people who are not from Estonia or not a resident of Estonia a door to enter services like business ownership, digital contracts signing, banking, taxing, payment processing and notary services.

%\subsection{Uport} 
Uport \cite{lundkvist2017uport} is a self-sovereign identity platform based on public Ethereum Blockchain. 
Like Sovrin, Uport supports attribute based data sharing, decentralized identifiers and verifiable credentials, and it does not store any private data on the public ledger. However, as oppose to our system, it uses a public Blockchain, and a cost is associated with transactions in the network.

\section{Conclusion}
\label{conclusion}
In this study, we focused on laying a foundation for ``decentralized digital identity'' in Blockchain as ``self-sovereign digital identity'', supported by modern cryptography and verifiable digital credentials. We described the problems and challenges exist in traditional identity management methods in terms of security, privacy, usability and globalization. 
We reviewed existing solutions in the literature, and proposed a system which leverages powerful features of Blockchain to realize a true private, secure and globally usable digital identity solution, in which identity owners fully own and control their portable identity and identity based records without depending on centralized authorities. For future work, we intend to explore possibilities of integrating our solution on mobile applications, and creating a crypto-currency to fuel the incentive of consent based data sharing through the concept of verifiable credentials.

%\section{Section in Appendix}
%\label{appendix-sec1}

%% References
%%  \cite{key}        ==>>  [#]


%\bibliographystyle{elsarticle-num}
%\bibliographystyle{elsarticle-harv}
% \bibliographystyle{elsarticle-num-names}
% \bibliographystyle{model1a-num-names}
% \bibliographystyle{model1b-num-names}
% \bibliographystyle{model1c-num-names}
% \bibliographystyle{model2-names}
% \bibliographystyle{model3a-num-names}
% \bibliographystyle{model3-num-names}
% \bibliographystyle{model4-names}
% \bibliographystyle{model5-names}
% \bibliographystyle{model6-num-names}

\bibliographystyle{apalike}
\bibliography{Sample}

\end{document}